\documentclass[aps,pre,twocolumn,showpacs,amssymb]{revtex4}
\usepackage{blindtext} 
\usepackage{amsmath,bm,amssymb,graphics,epsfig}
\usepackage{amsmath,graphics,epsfig} \usepackage{multirow}
\usepackage{supertabular} \usepackage{array} \usepackage{longtable}
   
\newcommand{\gae}{\hbox{\lower0.7ex\hbox{$\sim$}\llap{\raise0.4ex\hbox{$>$}}}}
\newcommand{\lae}{\hbox{\lower0.7ex\hbox{$\sim$}\llap{\raise0.4ex\hbox{$<$}}}}
 
\makeatletter
\newcommand*{\rom}[1]{\expandafter\@slowromancap\romannumeral #1@}
\renewenvironment{widetext@grid}{%
 \par\ignorespaces
\setbox\widetext@top\vbox{%
\vskip15\p@
\hb@xt@\hsize{%
\leaders\hrule\hfil
\vrule\@height6\p@
}%
\vskip6\p@
 }%
\setbox\widetext@bot\hb@xt@\hsize{%
\vrule\@depth6\p@
\leaders\hrule\hfil
}%
\onecolumngrid
 \let\set@footnotewidth\set@footnotewidth@ii
 }{%
\par
\twocolumngrid\global\@ignoretrue
\@endpetrue
}%
\makeatother
\begin{document}
\title{Equivalent-neighbor percolation models in two dimensions: 
crossover between mean-field and short-range behavior} 
\author{Yunqing Ouyang~$^{1}$~\footnote{email: oymj@mail.ustc.edu.cn},
Youjin Deng~$^{1}$~\footnote{Corresponding author; email:
yjdeng@ustc.edu.cn}, 
Henk W.~J. Bl\"ote~$^{2}$ }
\affiliation{$^{1}$ Hefei National Laboratory for Physical Sciences at
Microscale, Department of Modern Physics, University of Science and Technology
of China, Hefei 230027, China }
\affiliation{$^2$Lorentz Institute, Leiden University,  P.O. Box 9506,
2300 RA Leiden, The Netherlands} 
\date{\today} 
\begin{abstract}
We investigate the influence of the range of interactions in the
two-dimensional bond percolation model, by means of Monte Carlo simulations. 
We locate the phase transitions for several interaction ranges, as
expressed by the number $z$ of equivalent neighbors. We also consider
the $z \to \infty$ limit, i.e., the complete graph case, where percolation
bonds are allowed between each pair of sites, and the model becomes
mean-field-like.
All investigated models with finite $z$ are found to
belong to the short-range universality class.
There is no evidence of a tricritical point separating
the short-range and long-range behavior, such as is known 
to occur for $q=3$ and $q=4$ Potts models.
We determine the renormalization exponent describing a finite-range
perturbation at the mean-field limit as $y_r \approx 2/3$. 
Its relevance confirms the continuous crossover from mean-field
percolation universality to short-range percolation universality. 
For finite interaction ranges, we find approximate relations between
the coordination numbers and the amplitudes of the leading correction
terms as found in the finite-size scaling analysis.
\end{abstract}
\maketitle

\section {Introduction}
The range of the interactions plays an important role in phase transitions.
Systems with pair interactions decaying with a negative power of the 
distance are found to display a variety of universality classes,
depending on that power as well as on the dimensionality \cite{FMN,Sak,LB1}.
A different
way to modify the range of the interactions is specified in the so-called
equivalent-neighbor models, in which the pair interactions are constant
up to a range $r$ and zero at larger distances. These models are referred
to as medium-range models. The effects of such interactions were already
investigated in several two-dimensional model systems, including the Ising
model \cite{LBB}, i.e., the $q=2$ Potts model, and the $q=3$ and $q=4$
Potts models \cite{QDLGB}.  In these models, the range $r$ contributes
markedly to the irrelevant temperature field near the short-range
critical fixed point, and thereby to the corrections to scaling.

In the limit of $r\rightarrow \infty$, the
equivalent-neighbor models become mean-field-like, while for sufficiently
small $r$ they fall in the short-range universality classes.
Two different crossover scenarios are known between these two extremes.
In the Ising case, the crossover along the critical line is uniform from
the unstable mean-field
(MF) limit to the short-range fixed point \cite{LBB}, so that all models
with finite $r$ belong to the short-range universality class. 
In contrast, the MF fixed point is stable in the cases of the
$q=3$ and $q=4$ Potts models. For $q=3$ there is an intermediate
tricritical point \cite{QDLGB}, while for $q \to 4$ the critical and
tricritical points merge into a special fixed point \cite{NBRS} with
a marginal operator.

In order to provide a quantitative analysis concerning the scenario for
the percolation model, we investigate the equivalent-neighbor version 
of this model with a finite and variable interaction range $r$.
This model is defined by the probability distribution of the percolation
bonds, described by the partition sum
\begin{equation}
Z_{\rm perc} = 
\sum_{\mathcal G} p^{N_{\rm b}} (1-p)^{N_{\rm e}-N_{\rm b}} \,,
\label{Zperc}
\end{equation}
where the sum is on all graphs $G$ covering a number of edges of the square
lattice. This number ``bonds'' is denoted $N_{\rm b}$ and can assume all
values from 0 up to complete covering.  Edges connect each site to all of
its neighbors within the interaction range $r$.
The interaction range is roughly specified by $z \approx \pi r^2$, 
where the coordination number $z$ is the number of equivalent-neighbors
interacting with a site of the square lattice. Each edge has a probability
$p$ to be covered by ${\mathcal G}$. The total number of edges,
including those that are not occupied, is $N_{\rm e}$ . 

The analysis of the percolation model makes use of the connection between the
percolation model and the $q=1$ Potts \cite{Potts} model. To illustrate the
equivalence, we consider the partition sum of the $q$-state Potts model
\begin{equation}
Z_{\rm Potts} =\sum_{s_i} e^{K\sum_{\langle i, j
\rangle}\delta_{s_{i} s_{j}}} \,,
\end{equation}
where the $s_{i}$ sit on a
square lattice and assume the values $s_{i} = 1,2,\cdots,q$. At high
temperatures $T$ ($K$ small), the spins tend to take random values and the
model is disordered, whereas at low $T$ ($K$ large, $> 0$) the model is
ferromagnetic, the spins tend to be in the same state, even at large
distances.
\par Kasteleyn and Fortuin \cite{KF} introduced bond variables between
interacting neighbors, which can have two values: absent or present.
Then it is possible to sum out the Potts variables in $Z_{\rm Potts}$ and
one obtains \cite{KF}
\begin{equation}
Z_{\rm Potts} =
Z_{\rm RC} = \sum_{\mathcal G}u^{N_{\rm b}}q^{N_{\rm c}} \,,
\label{Zrc}
\end{equation}
where $u = e^{K}-1$, and the graph ${\mathcal G}$ represents the
Kasteleyn-Fortuin bond variables. The sum is on all such graphs,
as in the percolation model Eq.~(\ref{Zperc}). The number of connected
components (clusters) in ${\mathcal G}$ is denoted by $N_{\rm c}$.
In the random-cluster representation \cite{KF} of the Potts model,
the bond percolation model is obtained by taking the $q \to 1$ limit. 
Dividing out a trivial factor in Eq.~(\ref{Zrc}) then leads to 
\begin{equation}
e^{-KN_{\rm e}} Z_{\rm RC}=Z_{\rm perc} \,,
\end{equation}
with $Z_{\rm perc}$ as given by Eq.~(\ref{Zperc}).

We are now facing the question whether, for $q=1$, the MF to
short-range crossover is uniform, like for $q=2$, or more like the
$q>2$ Potts models.
We investigate this question by simulating systems on $L\times L$ square
lattices, for a sequence of finite sizes $L$ with periodic boundary
conditions, by means of Monte Carlo 
method \cite{LB} which remains efficient 
for systems with interactions of a long range. We sampled the size of
the largest cluster, and the second and fourth moments of the cluster-size
distribution. Thus we obtained the percolation equivalent of the Potts
magnetization moments and the Binder ratio, a dimensionless number related
to the Binder cumulant \cite{Bi},  as explained in Sec.~\ref{secint}.
We also sampled the wrapping probabilities, also 
described in Sec.~\ref{secint}. For $z=4$, the nearest-neighbor model, the
critical point \cite{essam} and the universality class \cite{CG} are known.
To achieve our analysis of the crossover between MF and short-range
percolation, we stepwise increase the coordination number from $z=4$ 
towards infinity, meaning that each site can interact with all other
sites, the mean-field or complete-graph case. 

For each of these values $z>4$, the percolation threshold is determined
from the simulation results.
At the same time, we investigate whether the critical behavior still
belongs to the short-range universality class. For this purpose it is
necessary to simulate systems of sizes that are sufficiently large  in
comparison with the finite range of the interactions.
We apply a finite-size scaling (FSS) analysis to a few different 
observables, in order to determine the nature of the crossover phenomena.  
Since the amplitudes of the corrections to scaling are a measure of the 
irrelevant fields, we may expect that the finite-size scaling analysis
will provide some useful information.

The outline of the rest of this paper is as follows.
In Sec.~\ref{secint} we provide an overview of some known aspects of the
two-dimensional percolation model, including the MF description.
This section also describes the role of magnetic quantities, and 
defines the Binder ratio defined on the critical cluster-size distribution.
It also includes the finite-size scaling of the Binder ratio 
and of the critical wrapping probabilities,
taking into account the logarithmic factors in the corrections to
scaling, which are caused by the presence of  two degenerate irrelevant
exponents.
Section \ref{secmet} explains the Monte Carlo simulation method, 
and describes the sampled quantities.
In Sec.~\ref{secres} we present the results of the analysis of the wrapping 
probabilities, for the models with $ 4 \leq z \leq 60 $. We verify the
universality of these models, and determine the amplitudes of the
leading corrections to scaling. Details of this analysis are given in
Appendix \ref{coramp}, together with an investigation of several models
that do not obey the condition stated under Eq.~(\ref{Zperc}), namely
that the interacting neighbor sites fill a circle of a given radius $r$.
Thus one can purportedly distinguish the effects due the number $z$ of
neighbors and their distances on the correction amplitudes.

Results for the Binder ratio are included in Sec.~\ref{secres}, for
models with finite $z$ as well as in the MF limit $z \to \infty$.
Special attention is given to the crossover between the MF
fixed point and the short-range percolation model, for the case of the
Binder ratio as well as for the temperature and magnetic exponents.
Finally, Sec.~\ref{secres} includes a numerical determination of the
crossover exponent of the MF model associated with a finite interaction
range.  The paper concludes with a discussion of the main results in
Sec.~\ref{seccon}.

\section{Existing theory and results}
\label{secint} 
\subsection{Percolation and magnetism }
We illustrate the role of the magnetism in percolation, using the
context of the $q$-state Potts model. The Hamiltonian for that model,
including an external field is
\begin{equation}
 {\mathcal H} = -K \sum_{<i,j>} \delta(\sigma_i, \sigma_j)
- H \sum_{k=1}^N \delta(\sigma_k,1) \,,
\label{potham}
\end{equation}
where  the magnetic field $H$ acts on Potts state 1 only.
The interaction in Eq.~(\ref{potham}) can be alternatively
formulated in terms of $(q-1)$-dimensional Potts spin vectors,
in order to reflect the full geometric symmetry.
This is achieved by substituting
\begin{equation}
\delta(\sigma_i,\sigma_j) =
\frac{1}{q}[1+(q-1)\vec{e_{\sigma_i}}\boldsymbol{\cdot} \vec{e_{\sigma_j}}] \,,
\end{equation}
where $\vec{e_1},\vec{e_2}, \cdots,\vec{e_q}$ are $q$ unit vectors pointing
from the center to the corners of a regular hypertetrahedron in $q-1$
dimensions. Thus
\begin{equation}
\vec{e_{\sigma}}\boldsymbol{\cdot}\vec{e_{\sigma'}} =
\frac{q \delta(\sigma, \sigma')-1}{q-1} \, .
\label{psprod}
\end{equation}
The magnetization density of an $N$-spin system is
\begin{equation}
\vec{m}\equiv \sum_{i=1,N} \frac{\vec{e_{{\sigma}_i}}}{N} \,.
\end{equation}
The magnetization of the system along direction 1 contains a contribution
from the spins in state 1, as well as contributions from the other
$q-1$ states which are, according to Eq.~(\ref{psprod}), weighted with 
a factor $-1/(q-1)$. This still allows for a zero `magnetization' of the
$q \to 1$ random-cluster model in its the disordered phase.
Since all Potts spins in a cluster are in the same state, and 
different clusters are in randomly distributed Potts states, the size
of the largest cluster is a good measure of the Potts spontaneous
magnetization.

Furthermore, the second and fourth magnetization moments can be expressed
in terms of the random-cluster size distribution as
\begin{equation}
\langle m^2\rangle =\langle \sum_i c_i^2 \rangle\,,
\label{m2}
\end{equation}
and
\begin{equation}
\langle m^4\rangle=\left\langle\frac{q+1}{q-1}\left(\sum_i c_i^2\right)^2 -
\frac{2}{q-1} \sum_i c_i^4\right\rangle \, ,
\label{m4}
\end{equation}
where $c_{\rm i}$ is the density of cluster $i$. Equations~(\ref{m2}) and
(\ref{m4}) enable the sampling of magnetic quantities in the Potts and
random-cluster models without using actual Potts spins. While sampling
of Eq.~(\ref{m4}) for $q=1$ is obviously problematic, Eqs.~(\ref{m2})
and (\ref{m4}) indicate that even  powers of the magnetization scale as
sums over the same powers of the cluster sizes.

\subsection{Finite-size scaling}
\par To provide a basis for our finite-size scaling analysis,
we write down the scaling equation for the free energy density of a model
as a function of the finite size $L$ and the scaling fields, namely
the relevant temperature field $t$, the relevant magnetic field $h$,
and two irrelevant fields $u_1$ and $u_2$:
\begin{displaymath}
f(t, h, u_1, u_2,L)=f_{\rm a}(t, h, u_1, u_2,L)
\end{displaymath}
\begin{equation}
+b^{-d}f_{\rm s}( b^{y_t}t, b^{y_ h}h, b^{y_1}u_1,
b^{y_2}u_2,L/b)  \,,
\label{fscal}
\end{equation}
where $f_{\rm a}( t,h, u_1, u_2,L)$ is
the analytic part of the free energy and $f_{\rm s}(t, h, u_1, u_2,L)$
is the singular part, and $b$ is the linear scale factor of the transformation. 
The prefactor $b^{-d}$ is due to rescaling of the $d$-dimensional volume.
The relevance of $t$ and $h$ means that the associated exponents $y_t$ and
$y_h$ are positive, while the irrelevant exponents $y_1$ and $y_2$ are negative.
In the present percolation models, the distance $p_{\rm c}-p$ plays,
up to a model-dependent multiplicative constant, the role of the
temperature field $t$. The substitution $b = L$ leads to 
the FSS relation for the free energy:
\begin{displaymath}
f(t, h, u_1, u_2,L)= f_{\rm a}(t,h, u_1, u_2,L)+
\end{displaymath}
\begin{equation}
L^{-d} f_{\rm s}
( L^{y_t} t ,L^{y_h}h, L^{y_1}u_1, L^{y_2}u_2,1)  \,.
\label{fsscal}
\end{equation}
The Binder ratio $Q$ is defined as
\begin{equation}
Q(t,L)=\langle m^2\rangle^2 / \langle m^4\rangle  \,,
\label{Qdefm}
\end{equation}
where $\langle m^2 \rangle$ and $\langle m^4 \rangle$ are the magnetization
moments of order two and four respectively:
\begin{equation}
\langle m^2 \rangle  =
L^{-d}\left(\frac{\partial^2f}{\partial H^2}\right)_{H = 0} 
\end{equation}
\begin{equation}
\langle m^4 \rangle =
L^{-3d}\left(\frac{\partial^4 f}{\partial H^4}\right)_{H=0}+
3L^{-2d}\left(\frac{\partial^2f}{\partial H^2}\right)^2_{H=0} \,,
\end{equation}
where $H$ is the physical magnetic field.
The correspondence between the derivatives with respect to $h$ and $H$ is 
\begin{equation}
\frac{\partial^2 f}{\partial H^2} =
f^{(2)}\left(\frac{\partial h}{\partial H}\right)^2 +\cdots \,,
\label{d2fdH2}
\end{equation}
and
\begin{equation}
\frac{\partial^4 f}{\partial H^4}
=f^{(4)}\left(\frac{\partial h}{\partial H}\right)^4+
2f^{(2)}\frac{\partial h}{\partial H}\frac{\partial ^3 h}{\partial H^3}
+\cdots  \,.
\end{equation}
For the present we neglect additional terms due to the dependence of the
other scaling fields on $H$.
Furthermore, by differentiating $k$ times with respect to $h$, one obtains
\begin{displaymath}
f^{(k)}(t, h, u_1, u_2,L) =
\end{displaymath}
\begin{displaymath}
L^{ky_h-d}
f^{(k)}_{\rm s}(L^{y_t}t, L^{y_h}h, L^{y_1}u_1, L^{y_2}u_2,1)+
\end{displaymath}
\begin{equation}
f_{\rm a}^{(k)}(t, h, u_1, u_2,L) \,,
\end{equation}
and $\langle m^2 \rangle$ can be explicitly expressed as
\begin{equation}
\langle m^2 \rangle = L^{2y_h-2d}\left(\frac{\partial h}{\partial H}\right)^2
(f_{\rm s}^{(2)} +L^{d-2y_h}f_{\rm a}^{(2)})+\cdots \,.
\label{m2sc}
\end{equation}
The leading term in $\langle m^2\rangle^2$ is of order $L^{4y_h-4d}$, as
well as that in $\langle m^4 \rangle$. The $L^{4y_h-4d}$ factors in $Q$
due to $\langle m^2\rangle^2$ and $\langle m^4 \rangle$ cancel, and the
$L^{d-2y_h}f_{\rm a}^{(2)}$ term in $\langle m^2 \rangle$
indicates that there should be a correction term in $Q$ due to the
analytic background that behaves as $L^{d-2y_h}$.
Furthermore, the nonuniversal geometric factors $(\partial h/\partial H)^4$
in the numerator and denominator of $Q$ cancel, and $f_{\rm s}$ is a
universal function, so that it follows that the resulting
finite-size scaling equation for $Q$ is universal in the scaling limit.

Next, we return to Eq.~(\ref{d2fdH2}) in which we have neglected a term
$(\partial f/\partial t) (\partial^2 t/\partial H^2)$.
In general one expects that the temperature scaling field includes a
quadratic term in the physical magnetic field $H$:
\begin{equation}
t= \sum_{k} \tau_k (p-p_{\rm c})^k + \rho H^2 +\cdots\,. 
\label{tsc}
\end{equation}
As a consequence, the $t$-dependence of Eq.~(\ref{fsscal}) leads to an
additional correction term proportional to $L^{y_t-2d}$
in Eq.~(\ref{m2sc}). The relative scale of this correction
is $L^{y_t-2y_h}$. It similarly affects $\langle m^4 \rangle$ and $Q$. 
Combining the universal part and the additional corrections, we obtain
\begin{displaymath}
Q(t,u_1, u_2,L)=
\end{displaymath}
\begin{equation}
Q(L^{y_t}t, L^{y_1}u_1,L^{y_2}u_2,1)+b_3L^{d-2y_h}+
b_4L^{y_t-2y_h} + \cdots\,.
\label{Qsc}
\end{equation}

The analytic part of the free energy does not contribute to the
wrapping probabilities $R_{\rm w}$, which are, like $Q$, dimensionless.
The finite-size scaling formula is
\begin{equation}
R_{\rm w}(t,u_1,u_2,L )=
R_{\rm w}(L^{y_t}t, L^{y_1}u_1, L^{y_2}u_2,1)  \,.
\label{Rsc}
\end{equation}

\subsubsection{Degenerate irrelevant exponents}
\label{dirrex}
\par 
In the $d=2$ Potts model, there is an exponent $-$2 for general $q$. For a
numerical justification, see \cite{XF}. In addition, the second thermal
exponent of the Potts model depends on $q$ and assumes the value $-2$
in the percolation case $q=1$ \cite{CG}. Under these circumstances, the
the irrelevant exponents become degenerate and the corresponding
scaling fields could couple with each other \cite{Wegner}. Suppose
that, in differential form, one has the following scaling equations for
the irrelevant fields 
\begin{eqnarray}
\frac{du_{\rm i}(l)}{dl} &=& y_{\rm i}u_{\rm i}(l) \\
\frac{du_{\rm l}(l)}{dl} &=& y_{\rm i}(u_{\rm l}(l)+
\alpha u_{\rm i}(l)) \,,
\end{eqnarray}
where $l$ parametrizes the renormalization flow
such that the linear scale factor is $b = e^{l}$. The solution is
\cite{Wegner} 
\begin{eqnarray}
u_{\rm i}(l) &=& b^{y_{\rm i}}u_{\rm i}(0)\\
u_{\rm l}(l) &=& b^{ y_{\rm i}}(u_{\rm l}(0) +\alpha u_{\rm i} \ln(b)) \,,
\end{eqnarray}
which shows that the rescaling of the irrelevant fields as used in
Eqs.~(\ref{Qsc}) and (\ref{Rsc}) has to be modified by adding a logarithmic
factor.
After substitution of the correct rescaling of $u_{\rm i}$ and 
$u_{\rm l}$ into the scaling equations Eqs.~(\ref{Qsc}) and Eqs.~(\ref{Rsc}),
setting $b=L$, and Taylor expansion, one obtains
\begin{eqnarray}
Q(t,L) &=& Q+\sum_{i=1}^{\infty}q_{i}t^{i}L^{i\textit{y}_t}+
b_{\rm i}L^{y_{\rm i}}+ \nonumber \\
b_{\rm l}\ln(L)L^{y_{\rm i}} &+& b_3L^{d-2y_h} +b_4L^{y_t-2y_h}+ \cdots
\label{fsq}\\
R_{\rm w}(t,L) &=& R_{\rm w}+\sum_{i=1}^{\infty}a_{i}t^{i}L^{i\textit{y}_t}+
b_{\rm i}L^{y_{\rm i}}+ \nonumber \\
 &  & b_{\rm l}\ln(L)L^{y_{\rm i}} +\cdots \,.
\label{fsr}
\end{eqnarray}
These formulas describe the
deviations from universal values $Q$ of the Binder ratio and $R_{\rm w}$
of the wrapping probabilities in finite systems near the critical point,
thus providing the basis for the FSS analysis of the numerical data.

Since, close to the critical point, $p-p_{\rm c}\propto t$, Eqs.~(\ref{fsq})
and (\ref{fsr}) describe the dependence of these two dimensionless
quantities on the bond probability $p$ and system size $L$.
The polynomial function in $p$
takes into account the effect caused by the deviation of $p$ from the
critical bond probability $p_{\rm c}$, while the critical finite-size effect
is described by the correction terms
$b_{\rm i}L^{y_{\rm i}}+b_{\rm l}\ln(L)L^{y_{\rm i}}$ due to irrelevant
scaling fields. Both $b_{\rm i}$ and $b_{\rm l}$ are analytic 
functions of $u_1$ and $u_2$.
In addition there is an analytic background term with amplitude $b_3$ for
the Binder ratio.

For models with small coordination
number $z$, deviations from isotropy are relatively large
and will be taken into
consideration, which could be observable in the correction
terms, $b_{\rm i}L^{-2}$ and $b_{\rm l}\ln(L)L^{-2}$.  
\par 
For percolation in two dimensions, several critical exponents are 
exactly known \cite{CG}. The leading and subleading thermal
renormalization exponents are $y_t = 1/\nu = 3/4$, and $y_i
= -2 $ respectively; the leading critical exponent in the
magnetic sector is $y_h = 91/48$, and the exponent of the correction 
due to the analytic background term is thus $-43/24$.

\subsection{Mean-field percolation}
\label{mfpxp}
\par In mean-field (MF) percolation, every site can connect to any other
site.  The behavior of this model is well known  \cite{MS,WFY,Dro,Bol}.
We summarize the derivation of the temperature and magnetic exponents.
These exponents do also apply in high-dimensional systems \cite{KB} with
short-range couplings. In the MF case, one has $z=L^2$,
and the  edges of the lattice thus form a complete graph.
The critical bond probability $p$ is thus inversely related to the total
number of sites $L^2$. It makes therefore sense to work with the
integrated bond probability $p_{i} \equiv p L^2$.

To construct the self-consistency equation for this model, we again use the
framework of the $q$-state Potts model.  The Hamiltonian for this model 
on the complete graph is, analogous to Eq.~(\ref{potham}),
\begin{equation}
 {\mathcal H} = -\frac{K}{N}\sum_{i< j}^N \delta(\sigma_i, \sigma_j)
- H \sum_{k=1}^N \delta(\sigma_k,1) \,.
\label{hamilton}
\end{equation}

We expect that, at high temperatures, i.e., small $K$, $\vec{m}=0$ in the
thermodynamic limit, and that, at low temperatures, a spontaneous
magnetization exists in the direction of one of the unit vectors
mentioned, for example, direction 1. We thus formulate our analysis in
terms of that component $m=|\vec{m}|$ of the magnetization density.

The canonical probability $p_1$ that a spin assumes state 1 is
\begin{equation}
p_1=\frac{e^{KN_1/N+H}}{e^{KN_1/N+H}+(q-1)e^{K(N-N_1)/[N(q-1)]}}\,,
\end{equation}
where $N_1/N=[1+(q-1)m]/q$ is the density of spins in state 1.
Substitution of $N_1$ yields
\begin{equation}
p_1=\frac{e^{Km+H}}{e^{Km+H}+q-1}\,.
\end{equation}
Expansion of this expression to first order in $H$ and second order in
$m$ leads to
\begin{equation}
p_1=\frac{1}{q} +(q-1)\left[
\frac{Km+H}{q}-\frac{K^2m^2(2-q)}{2q}\right] \,.
\label{p1}
\end{equation}
The requirement of self-consistency imposes that $m=(qp_1-1)/(q-1)$ or
\begin{equation}
p_1=\frac{1}{q} +m\frac{q-1}{q} \,.
\end{equation}
Substitution in Eq.~(\ref{p1})
leads to a quadratic equation in $m$
\begin{equation}
\frac{(1-t)^2(2-q)m^2}{q} +2tm -2H =0 \,,
\label{sce1}
\end{equation}
where we have substituted the temperature-like variable $t\equiv 1-K$.
We have divided out a factor $q-1$, and the interpretation of
this equation for $q=1$ has to be done with some care. The variable
$m$ loses its meaning as a magnetization, but it keeps its meaning as
the density of the largest cluster in the ordered phase, also in
the $q \to 1$ limit of the random-cluster model.
In this limit, Eq.~(\ref{sce1}) reduces, for small $t$, to
\begin{equation}
m^2 +2tm -2H =0 \,,
\label{sce2}
\end{equation}
which provides a description of the critical region.
 The critical point is located at $t=H=0$.  For $H=0$, one has $m \propto -t$,
which determines the critical exponent $\beta=1$. For  $t=0$, one has
$m \propto \sqrt H$, so that $\delta=2$. Differentiation to $H$ yields
the susceptibility-like quantity $\chi \propto t^{-1}$ at $H=0$,
and therefore $\gamma = 1$.

The exponents $\beta, \gamma, \delta$ can be expressed in effective
renormalization exponents and the dimensionality $d$ of the system as 
\begin{equation}
\beta=\frac{d- y_h^*}{y_t^*},~~~~\gamma=\frac{2 y_h^*
-d}{y_t^*} ,~~~~\delta=\frac{y_h^*}{d- y_h^*} \,,
\label{hyper}
\end{equation}
which yields 
\begin{equation}
y_h^*=2d/3 ,~~~~ y_t^*=d/3 \,.
\end{equation}
\par The asterisks in $y_{h}^*$ and $y_{t}^*$ may require some explanation.
The actual renormalization exponents $y_t$ and $y_h$ do not satisfy the
relation in Eqs.~(\ref{hyper}) without asterisks. The reason is that MF
models violate ``hyperscaling'' relations: scaling relations that involve
the dimensionality $d$  \cite{BNPY,BZJ}.  The renormalization exponents
with an asterisk describe the rescaling of the observables that are
conjugate to the scaling fields.  The notation with asterisks serves to
reproduce the usual dependence of the critical exponents on $y_t$ and $y_h$, 
as expressed by Eqs.~(\ref{hyper}).
In our case, $d = 2$, $y_h^* = \frac{4}{3}$, and $y_t^* = \frac{2}{3}$.
The critical exponent $\alpha = (2y_t^* - d)/y_t^* = -1$ describes a
specific-heat-like quantity in the percolation model \cite{HBZD}, and the
exponent $\beta$ describes the size of the largest percolation cluster
as a function of $p-p_{\rm c}$. The magnetic exponent  also determines the
finite-size scaling of the largest cluster at $t=0$ as asymptotically
proportional to $L^{y_h^*}$ \cite{Bol}.

\section {Methods}
\label{secmet}
\subsection{Simulation algorithms}
The simulation of models with many
interacting neighbors tends to be time consuming, if
all these bonds are individually taken into account.
This problem is circumvented by means of an
algorithm \cite{LB}, employed earlier in the
analysis of the medium-range Potts model with $q=2$
\cite{LBB}, and with $q=3$ and 4 \cite{QDLGB}. 
The computer time needed depends only weakly on the 
number of interactions per site.  The
application of this medium-range algorithm to the
percolation case $q=1$ is straightforward.

\subsection{Sampled quantities}

The wrapping probabilities $R_{\rm w}$ investigated in this paper are 
specified as $R_{\rm b}$, $R_{\rm e}$ and $R_1$, defined as follows.
The probability $R_1$ counts the events that a configuration of
percolation bonds connects to itself along the $x$ direction,
but not along the $y$ direction. For better statistical accuracy,
we count every configuration which percolates in one direction,
but not in both directions, and divide the total percolations by a
factor of 2; $R_{\rm b}$ is for simultaneous wrapping in both directions;
$R_{\rm e}$ is for wrapping in at least one direction \cite{Huhao}. 
The relevant factors that may be related to their universal values are
the number of couplings between a central site and its neighbors within
the interaction range, the boundary conditions and the shape of the system.
To what extent the microscopic couplings between the central site and 
its different neighbors, i.e., the interaction range, are important is 
the present subject of investigation.

In this analysis, it is very useful that the universal values of
the wrapping probabilities are known for models with short-range
interactions and periodic boundary conditions in a square geometry.
These are $R_{\rm b}=0.351642855$, $R_{\rm e}= 0.690473725$ and
$R_1=0.169415435$  \cite{ZLK}.

The Binder ratio $Q$ of the $q$-state Potts model is defined by
Eq.~(\ref{Qdefm}) on the basis of the distribution of the magnetization
density $m$. Using the random-cluster representation, the magnetization
moments, and thus $Q$, can be determined from the random-cluster size
distribution via Eqs.~(\ref{m2}) and (\ref{m4}). 
The latter quantity is divergent for $q \to 1$, but for a  scalable
cluster-size distribution, we may instead just define a universal ratio
for the percolation case by using the Ising-like definition with $q=2$:
\begin{equation}
Q=\frac{\langle \sum_i c_i^2 \rangle^2}
{\langle 3(\sum_i c_i^2)^2 -2\sum_i c_i^4\rangle} \,.
\label{Qdefc}
\end{equation}
This definition was used during the simulations.
Its universal value for two-dimensional percolation in a square
periodic geometry has been estimated as $Q=0.87048(5)$ \cite{QDB1,DB}.


\section {Results}
\label{secres}
\subsection{Analysis of the wrapping probabilities} 
\par We simulated finite systems with medium-range interaction,
and sampled the three kinds of wrapping
probabilities mentioned earlier, $R_{\rm b}, R_{\rm e}$ and $R_1$. 
In this work we restrict ourselves to systems with square symmetry,
so that we can analyze our data in the light of the known universal wrapping
probabilities \cite{ZLK} of the short-range model.
The coupling strengths for all the $z$ interacting neighbors are the same.

The simulations took place for roughly 30 different system sizes,
ranging from small ones, with $L$ slightly larger than the interaction
range, up to $L=2048$ in some cases. The larger systems serve to approximate
the asymptotic behavior in the thermodynamic limit, including the
determination of the critical point and the renormalization exponents
$y_t$ and $y_h$. The corrections to scaling in the $R_{\rm w}(p,L)$ data 
(w=b, e, or 1) are relatively small for large $L$, so that the corresponding
error sources are suppressed. 
On the other hand, relatively small systems are needed to obtain accurate
estimations of the amplitudes of $b_{\rm i}$ and $b_{\rm l}$ of the
corrections to scaling. The typical lengths of the simulations varied 
from millions of samples for the largest systems to billions for the
smallest ones.

First, we performed simulations of systems whose interaction regions 
are almost circular.  These interactions are fully determined by the
range $r$, such each site couples with all $z$ neighbors within that range.
The corresponding values of $z$ studied are $z=4$, 8, 12, 20,
28, 36, 48, 60, 224, 1224, and 4016. In Appendix \ref{coramp}, we shall
investigate the additional influence of the range of these interactions,
by relaxing the rule that the $z$ interacting neighbors fill a circle.
We also check for effects due to fourfold deviations 
of the sets of neighbors from isotropy.

\subsubsection{Tests of universality of the wrapping probabilities}
\label{testuwp}
\begin{figure} 
\begin{center} \leavevmode \epsfxsize 7.6cm \epsfbox{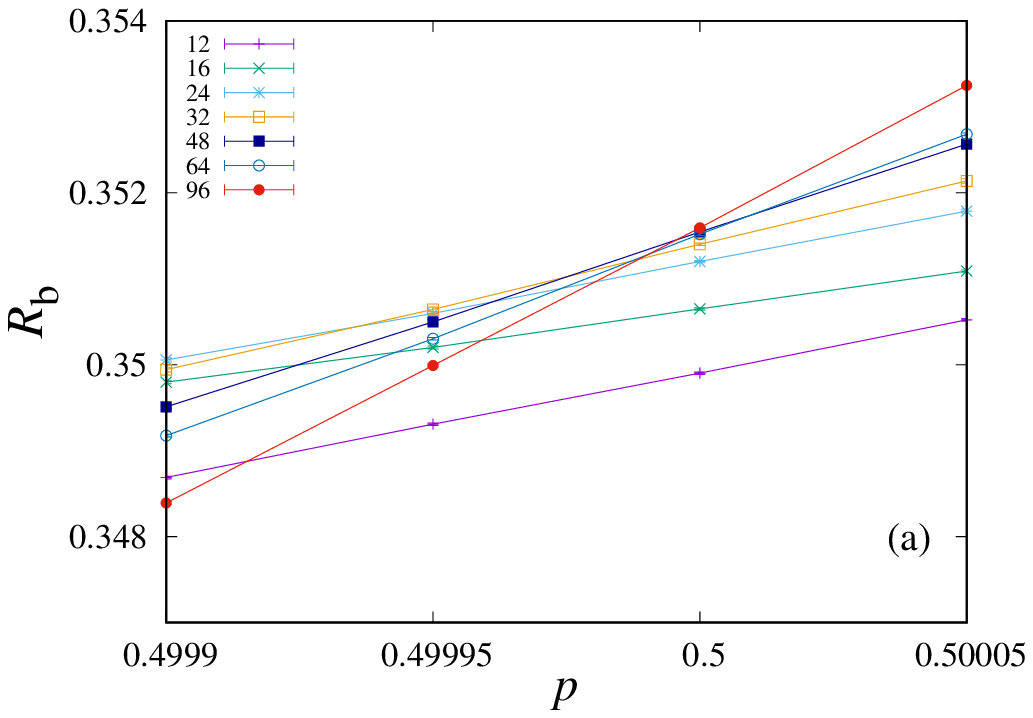}
\epsfxsize 7.6cm \epsfbox{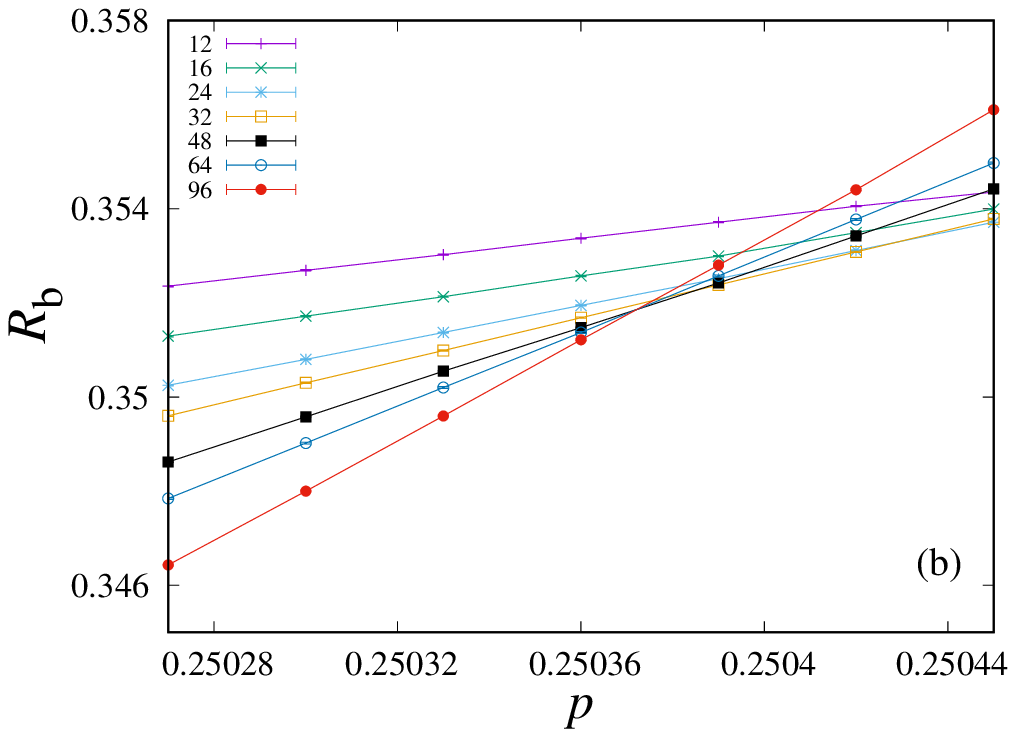} 
\end{center} 
\caption{$R_{\rm b}(L,p)$  versus $p$ plots for the 4- and 8-neighbor
models. Figure (a) applies to $z=4$ and (b) to $z=8$.
The system sizes are in the range from 12 to 96 as shown in the figure.
Larger systems correspond to steeper curves. } 
\label{rb48}
\end{figure}

\begin{figure}
\begin{center} \leavevmode \epsfxsize  7.6cm \epsfbox{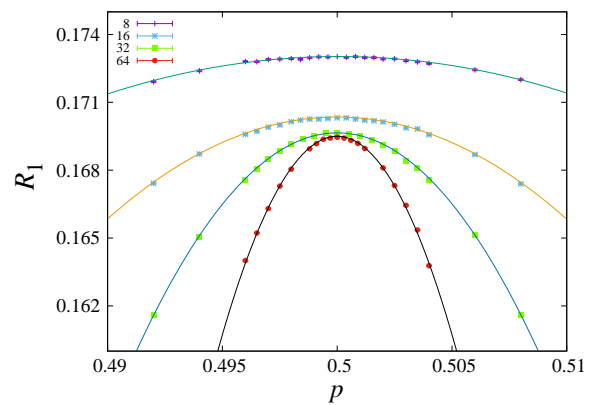}
\end{center}
\caption{$R_1(L)$ vs. $p$ for $L=8$, 16, 32 and 64 for the $z=4$ model.
Symmetry of this figure is imposed by the self-duality \cite{essam}
for $z=4$.}
\label{r14} 
\end{figure}

On the basis of Eq.~(\ref{fsr}) we expect that plots of the wrapping
probabilities versus $p$ for different $L$ display intersections,
at least if $a_1$ is nonzero.  For models in
the short-range percolation universality class, these intersections 
should, for large $L$, converge to the universal value of $R_{\rm w}$. 
The numerical results for the wrapping probabilities can thus be used
as a test whether models with $z>4$ do still belong to the short-range
universality class. This is illustrated in Fig.~\ref{rb48} for the case
of $R_{\rm b}$.
While the corresponding curves for $R_{\rm e}$ display a similar  behavior,
those for $R_1(p,L)$  appear to be different, because the amplitude $a_1$
is zero or very small. Self-duality imposes that $a_1=0$ for the $z=4$ model.
Thus, intersections are not useful for the determination of the critical point
from $R_1(p,L)$, as shown in Fig.~\ref{r14}. The amplitude of the quadratic
term is nonzero, as illustrated by the non-monotonic behavior of $R_1$. 
We fitted finite-size wrapping data by 
\begin{displaymath}
R_{\rm w}(p,L) = R_{\rm w} + \sum_k a_k(p-p_{\rm c})^kL^{ky_t}+
b_{\rm i}L^{-2}+
\end{displaymath}
\begin{equation}
b_{\rm l}\ln(L)L^{-2}+b_3L^{-3}\,.
\label{Rfit}
\end{equation}
which also takes into account 
a subleading correction of order $L^{-3}$.  Integer differences
between subleading exponents may be expected in general \cite{Cardy}.
Since the simulations were performed in a rather narrow neighborhood of
the critical point, higher-order terms in $p-p_{\rm c}$
rapidly become unimportant.

The fits for $R_{\rm b}$ and $R_{\rm e}$ yielded consistent estimates of
$p_{\rm c}$. These values of $p_{\rm c}$ were kept fixed in the fits
for $R_1$.
The resulting estimates for $R_{\rm b}$, $R_{\rm e}$ and $R_1$ are
in a fairly accurate agreement with the short-range universal
values, thus confirming that the investigated models with $z>4$ still
obey short-range universality. These results are listed in
Table \ref{uq}.

\begin{table}
\caption{Universal parameters as estimated for models with
different coordination numbers $z$.}
\vspace{0.5em}
\scalebox{1.0}{
{
\hspace*{-5mm} \begin{tabular}{||c|c|c|c|c|c||}
\hline
$z$ & $R_{\rm b}$ & $R_{\rm e}$ &     $R_1$    &  $y_t$    &  $y_h$     \\
\hline
4  &  0.351643(6) & 0.690464(5) & 0.169413(4)  & 0.749(2)  & 1.8958(2)  \\
8  &  0.351641(6) & 0.690479(5) & 0.169416(5)  & 0.748(2)  & 1.896(1)   \\
12 &  0.351649(4) &  0.69037(3) & 0.169411(3)  & 0.747(3)  & 1.895(2)   \\
20 &  0.35167(2)  &  0.69053(2) & 0.169409(3)  & 0.748(2)  & 1.894(3)   \\
28 &  0.35165(3)  &  0.69053(5) & 0.169407(5)  & 0.747(3)  & 1.895(2) \\
36 &  0.35163(3)  &  0.69051(3) & 0.169409(4)  &  0.748(3) & 1.897(2) \\
48 &   0.35167(3) &  0.69052(4) &  0.169405(4) &  0.746(4) & 1.896(1) \\
60 &   0.3517(1)  &   0.6904(1) &  0.16941(2)  &  0.747(2) & 1.895(2) \\
\hline
exact& 0.351642855 & 0.690473725  & 0.169415435  &  0.75   & 1.89583333 \\
\hline
\end{tabular}
}
}

\label{uq}
\end{table}
\subsubsection{Universality of the critical exponents }
\label{testu}
After confirming that the universal values of the wrapping probabilities of
$R_{\rm b}$ and $R_{\rm e}$ still apply for the values of $z$ investigated,
we fixed them in the fit formula while leaving $y_t$ as a free parameter
in Eq.~(\ref{fsr}). Again, the results are in agreement with the short-range
universal value $y_t=3/4$.  They are included in Table \ref{uq}.

As another test, we determine the magnetic renormalization exponent $y_h$.
We first fixed the wrapping probabilities and $y_t$ at
their universal values, and obtained improved estimates of the critical
points $p_{\rm c}$ of the models with $z>4$. The average density 
$c_{\rm l}$ of the largest cluster was then fitted by
\begin{displaymath}
c_{\rm l}(p,L) = L^{y_h-2}[\sum_{k=0,1,2,\cdots} g_k (p-p_{\rm c})^kL^{ky_t}+
\end{displaymath}
\begin{equation}
b_{\rm i}L^{-2}+b_{\rm l}\ln(L)L^{-2}+b_3L^{-3}+\cdots ] \,,
\label{clfit}
\end{equation}
with $p_{\rm c}$ as determined from the wrapping probabilities.
The results appear to be in a good agreement with the short-range
universal value $y_h=91/48$ \cite{CG}. The results are included in
Table \ref{uq}.
Finally, we performed a test if there is a contribution from the
second magnetic exponent $y_{h2}=19/48$ \cite{CG} to the wrapping
probabilities. If so, one may expect corrections with a relative
magnitude in the order $L^{y_{h2}-y_{h}}=L^{-3/2}$. However,
no evidence for such a correction was found.

\subsubsection{Determination of the correction amplitudes}
Having sufficient confidence in the validity of the short-range universal
parameters, we kept them fixed in the fits according to Eq.~(\ref{fsr}), 
with the purpose to find the best estimates of the critical points and the
amplitudes of the correction terms as a function of $z$.

\begin{figure}
\begin{center} \leavevmode \epsfxsize  7.6cm \epsfbox{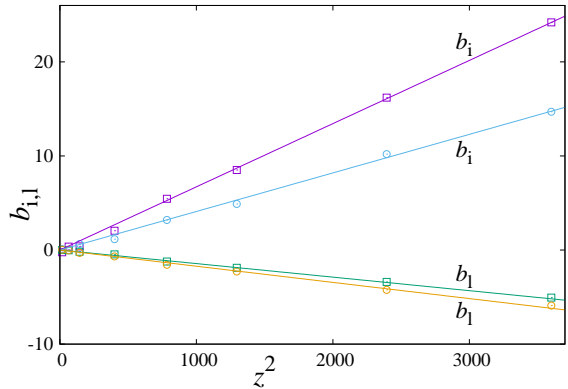}
\end{center}
\caption{Correction amplitudes $b_{\rm i}$ and  $b_{\rm l}$ in the
wrapping probabilities $R_{\rm b}(L,p)$ (squares) and $R_{\rm e}(L,p)$
(circles), as a function of $z^2$.
Data are shown for models with up to $z=60$ neighbors.  }
\label{bli}
\end{figure}

We find that the correction
amplitudes of the wrapping probabilities are approximately proportional
to $z^2$, and thus to the squared area within the interaction range.
This is illustrated in Fig.~\ref{bli} for the wrapping probabilities
$R_{\rm b}$ and $R_{\rm e}$.
The amplitudes $b_{\rm i}$ and $b_{\rm l}$, appear to become
small between $z=4$ and $z=8$,
suggestive of a short-range fixed point in that neighborhood. The
overall finite-size scaling effect due to the three correction terms
displays a change of sign between $z=4$ and 8, as can, for example,
be seen in Fig.~\ref{rb48}.
A more detailed analysis, including tabulated data for the correction
amplitudes and the critical points, is deferred to Appendix \ref{coramp}.

\subsection {FSS analysis for the Binder ratio $Q$} 
\subsubsection {Binder ratio of models with finite $z$}
\begin{figure} 
\begin{center} \leavevmode \epsfxsize  7.6cm \epsfbox{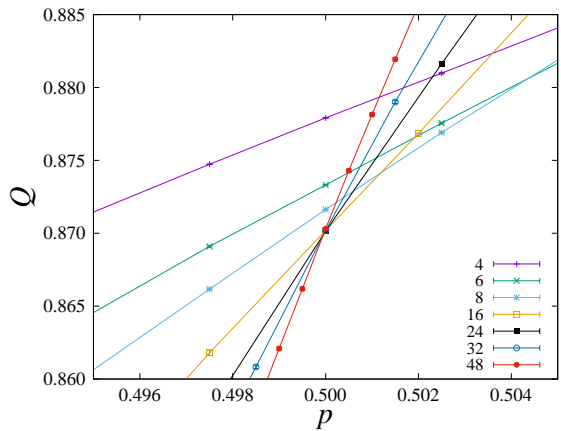}

\end{center} 
\caption{Binder ratio $Q$ of the bond-percolation model with
nearest-neighbor interaction vs. bond probability $p$, for several 
system sizes.  In the present $z=4$ case, the corrections are positive
for small $L$ and negative for large $L$.}
\label{q4n} 
\end{figure}

\begin{figure} 
\begin{center} \leavevmode \epsfxsize  8.0cm \epsfbox{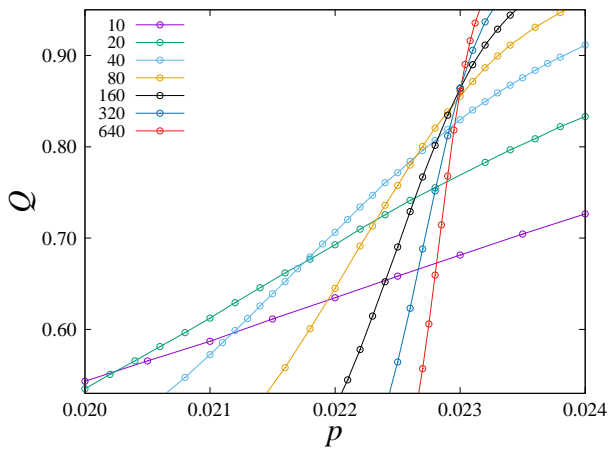}
\end{center} 
\caption{Binder ratio $Q(p,L)$ of the bond-percolation model with 60
equivalent neighbors vs.  bond probability $p$. The figure shows data
for several system sizes as indicated in the figure.  It displays
that strong corrections to scaling occur in the 60-neighbor model.}
\label{q60n} 
\end{figure}

The behavior of $Q(p,L)$, as defined by Eq.~(\ref{Qdefc}), of the
nearest-neighbor model ($z=4$) is illustrated in Fig.~\ref{q4n} for
several system sizes as a function of $p$. The intersections of the 
lines are seen
to converge to the critical point \cite{essam} at $p_{\rm c} = 0.5$. 
For small $L$, the combined finite-size corrections are positive at the 
critical point.
In contrast, they are negative for the $z \geq 8$ models, as seen e.g. in
Fig.~\ref{q60n}. The sign changes of the leading correction terms are
analogous to those of the wrapping probabilities.

Just as in the case of the wrapping probabilities, we analyzed several 
models with finite coordination numbers up to $z=60$. The increasing
distance to the short-range fixed point manifests itself as $z$ increases.
Figure \ref{q60n} shows that strong crossover effects already occur in
the 60-neighbor model:
the intersections between the curves with small $L$ are located at
relatively small values of $Q$, but for larger $L$ they still approach
the universal value \cite{QDB1,DB} of the short-range model.

\begin{figure} 
\begin{center} \leavevmode \epsfxsize 7.6cm \epsfbox{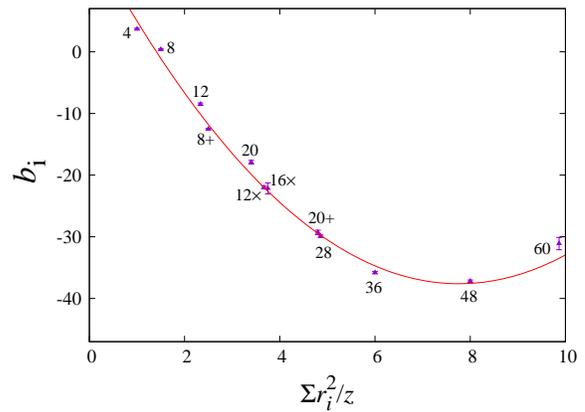}
\end{center} 
\caption{Correction to scaling amplitude $b_{\rm i}$ in the Binder ratio
$Q$ vs. $\sum_i r_i^2/z$. These data, which are taken from Table~\ref{qtab},
are approximately described by the parabola shown in the figure. 
} \label{bizq} 
\end{figure}

\begin{figure} 
\begin{center} \leavevmode \epsfxsize  8.0cm \epsfbox{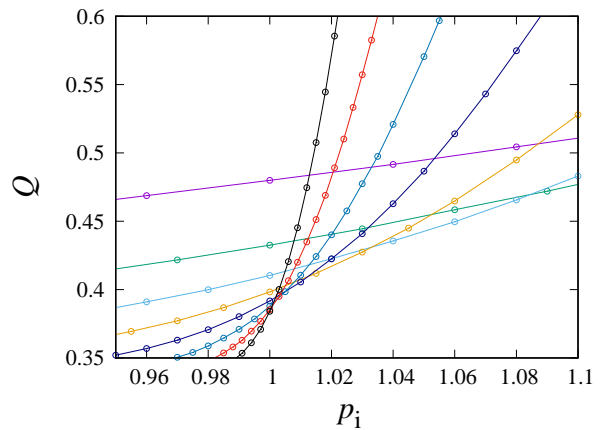}
\end{center} 
\caption{Binder ratio $Q(p,L)$ of the MF percolation model
vs. integrated  bond probability $p_{\rm i}$, for system sizes
$L=4$, 8, 16, 32, 64, 128, 256 and 512. Larger system sizes correspond
with steeper curves. The intersections of these curves approach 
$p_{\rm i} = 1$ for the larger systems.} 
\label{qmf}
\end{figure}

As described in Sec.~\ref{dirrex}, these $Q$ data are expected to
behave as
\begin{displaymath}
Q(p,L)=Q(p_{\rm c},\infty)+\sum_k q_k (p-1)^k L^{k y_t}+ b_{\rm i}L^{-2}+
\end{displaymath}
\begin{equation}
b_{\rm l}\ln(L)L^{-2}+b_3L^{d-2y_h}+b_4L^{y_t-2y_h}+\cdots\,.
\label{Qfit}
\end{equation}
For models in the short-range percolation universality class, we have
$y_t=3/4$ and $y_h=91/48$ \cite{CG}.
Since $p_{\rm c}$ can be determined more accurately from
the wrapping probabilities than from the Binder ratio, we fixed
the critical points at the values determined from the
wrapping probabilities. Least-squares fits were applied in rather
narrow $p$-intervals and included only few $p$-dependent terms.
In the case $z=4$, where the corrections appear to be small and we know the
exact critical point \cite{essam}, we could determine the universal value
of the Binder ratio as $Q=0.87057(2)$, slightly higher than a previous
result $Q=0.87048(5)$ \cite{QDB1,DB}. After checking that the Binder ratios
of the $z>4$ models were consistent with this value, the newer result for
$Q$ was used as a fixed parameter in the fits for the $z>4$ models.
The amplitudes  of the correction terms are found to increase rapidly
with $z$. This is illustrated  in Fig.~\ref{bizq}, which shows that
the relation between $b_{\rm i}$ and $z$ is approximately described by a
quadratic form, as shown by the parabola that figure.
The numerical results for the correction amplitudes in the Binder ratios
are listed in Appendix \ref{coramp}.

\subsubsection {Binder ratio of the complete graph}
As shown in Table~\ref{rb}, the critical point $p_{\rm c}$ is a decreasing
function of $z$.
In the limit $z \to \infty$ the bond percolation probability $p$ 
decreases as $1/z$. Since the MF solution of the percolation model
shows that $ zp_{\rm c} \to 1$ for $z \to \infty$,   
in simulations of finite systems we simply use a bond probability 
\begin{equation}
p=p_{\rm i}/L^2 
\end{equation}
so that  the
critical point of the integrated probability lies at $p_{\rm i}=1$.
Since the wrapping probabilities cease to be meaningful on the complete
graph, we restrict ourselves to the Binder ratio.
Results for $Q(p,L)$ of the complete graph are shown in Fig.~\ref{qmf}. 
As the system size $L$ goes up, the intersections of the $Q(p,L)$ curves 
are seen to converge to a value in agreement with the exact critical
point $p_{\rm i}=1$. The corresponding critical value of $Q$ is
clearly different from the universal short-range percolation value.

Fits based on the form of Eq.~(\ref{Qfit}) were applied, but using the
integrated bond probability $p_{\rm i}$ instead, with the critical point
fixed at $p_{\rm i,c}=1$. The leading correction exponent was found to
be close to $-2/3$, without a logarithm, in agreement with recent work of
Huang et al. \cite{Huang}. The $Q$ data are inconsistent with the
universality class of the finite-$z$ models.  After fixing $y_t=2/3$
and $y_h=4/3$, in accordance with Eqs.~(\ref{hyper}),
and leaving $Q$ as a free parameter, we obtained satisfactory fits.
The results, and further details, are included in Table \ref{qtab}.

\subsection{Crossover phenomena}
\subsubsection{Crossover of the Binder ratio }
\label{crBr}
\par
\begin{figure} 
\begin{center}
\leavevmode \epsfxsize 8.4cm \epsfbox{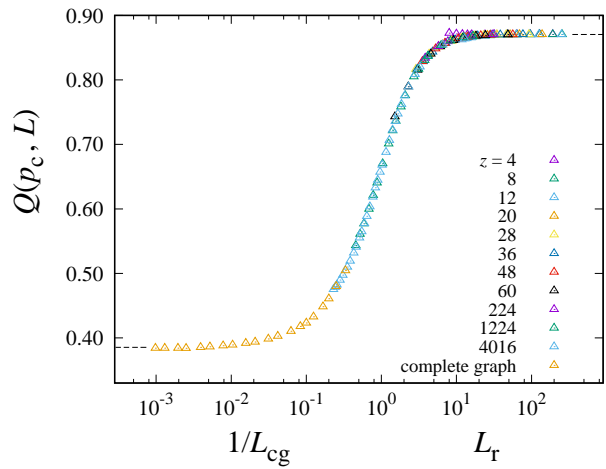} 
\end{center}
\caption{$Q(p_{\rm c},L)$ vs. rescaled size $L_{\rm r}$ for $z<\infty$.
The rescaling factor $b(z)$ in $L_{{\rm r}}$ is defined in the text
for the finite $z$ models.  The data for $z=4$ cannot be collapsed
similarly because the corrections have the opposite sign.
The figure includes data for the complete graph (cg), for which we adopt the 
inverse finite size $1/L_{\rm cg}$ as the finite-size parameter.
The asymptotic values of $Q$ for $L\rightarrow \infty$ are 0.87057 and
0.38126 respectively, marked by dashed lines in the figure.}
\label{crossover} 
\end{figure}

 
To demonstrate the crossover between the MF and short-range
models, we first construct a data collapse by means of a $z$-dependent
rescaling of the finite size. We define $L_{\rm r}\equiv L/b(z)$ with
$b(4)=b(8)=1$ and $b(z)$ for $z>8$ chosen such that the $Q(p_{\rm c},L)$
versus $L_{\rm r}$ curves collapse for $L\geq 8$. It is not possible to
include the $z=4$ data in this collapse because the corrections have the
opposite sign.
These results are shown in Fig.~\ref{crossover}. Also shown in this figure
is the finite-size dependence of the $Q$ results for the critical MF 
model versus the inverse finite size. Our use of $1/L_{\rm cg}$ and our
definition of $L_{\rm r}$ for the horizontal scale is tuned such that the 
finite-size data for $Q$ of the MF model just match with those of the
finite-$z$ models.

One observes that these data for $Q$ span the whole range between the
universal values of the MF and the short-range model. There is no 
sign of a possible intermediate multicritical point when $z$ increases.
The collapse of the other curves on that for $z=8$  involves a rescaling
factor $b(z)$ that satisfies the relation 
\begin{equation}
\begin{split}
b_{\rm i}(8)L^{-2}+b_{\rm l}(8)\ln(L)L^{-2}+b_3(8)L^{d-2y_h}=\mbox{\hspace{10mm}}\\ 
\mbox{\hspace{10mm}}b_{\rm i}(z)b(z)^2L^{-2}+b_{\rm l}(z)b(z)^2\ln[L/b(z)]L^{-2}+\\
\mbox{\hspace{10mm}}b_3(z)b(z)^{2y_h-d}L^{d-2y_h}\,.
\end{split}
\end{equation}
Among these correction terms, $b_3L^{d-2y_h}$ is the dominant one for
large $L$ and asymptotically determines $b(z)$ in that limit as
$b(z) \simeq [b_3(8)/b_3(z)]^{1/(2y_h-d)}$.  

\subsubsection{Crossover of the effective temperature exponent}

\par
\begin{figure}
\begin{center} \leavevmode \epsfxsize 7.60cm \epsfbox{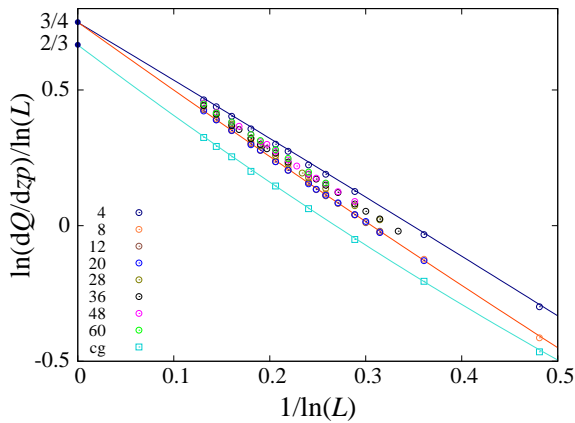}
\end{center}
\caption{Finite-size dependence of the derivative of the Binder ratio
$Q$ to the integrated bond probability $zp$. Different symbols indicate 
percolation models with different coordination numbers, as listed in
the figure. These results show an approximate data collapse for the
finite-$z$ systems.  There is a clear difference with the data for
the complete graph, or the MF limit, which are also shown.
The quantity plotted along the vertical scale is defined in the text,
and chosen such that the large-$L$ data should converge to the
temperature exponent $y_t$ which is 3/4 for short-range percolation,
and 2/3 for MF percolation. The data points for $z=4$, and $z=12$,
and for the MF model are connected by curves which are also intended
to guide the eye to the limiting value at $L = \infty$ on the vertical
scale.}
\label{lnQz}
\end{figure}

 As another demonstration of the crossover phenomenon, we analyze the
first derivative of $Q(p,L)$ with respect to the bond probability $p$ at
criticality \cite{QDLGB}.
From Eq.~(\ref{Qfit}) one expects that, 
at the transition point, it behaves as
\begin{equation}
\frac{\ln({\rm d}Q/{\rm d}p)}{\ln(L)} = y_{t} +
\frac{\ln(a_1)+\cdots}{\ln(L)} \,.
\end{equation}
The data for ${\rm d}Q/{\rm d}p$, were obtained from fits to the $Q$
versus $p$ simulation results and by numerical differentiation.
Since $p_{\rm c}$ is roughly proportional to $1/z$,
we include a factor $1/z$ in the derivative of $Q$ to $p$.
A comparison of the result, shown in Fig.~\ref{lnQz}, with similar
analyses for the $q=3$ and 4 Potts models in Ref.~\onlinecite{QDLGB}
illustrates the different dependences of these models on the range
of the interactions.

\begin{figure}
\begin{center} \leavevmode \epsfxsize 8.00cm \epsfbox{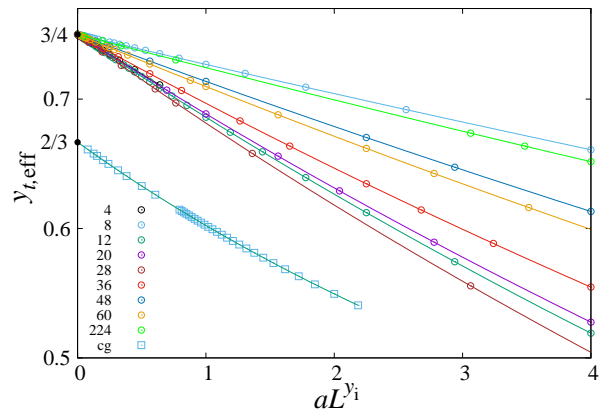}
\end{center}
\caption{Effective temperature exponent $y_{t,{\rm eff}}$, defined in
the text, as a function of $aL^{y_{\rm i}}$. In order to fit the
curves of the various interaction ranges in the same figure, the
amplitude $a$ is arbitrarily chosen as 8 for the mean-field limit,
and as $z^{-2}$ in the other cases. The horizontal scale used the
values $y_{\rm i}=-2/3$ and $-2$ respectively.
The data for each of the models show satisfactory large-$L$ convergence
to the expected values of the temperature exponent $y_t$, namely 3/4 for
the short-range models, and 2/3 for the MF case.  }
\label{ytef}
\end{figure}

The $z$-dependence of the effective temperature exponent can also be
illustrated with the quantity
\begin{equation}
y_{t,{\rm eff}}(L) \equiv
\ln [\{dQ(p,2L)/dp)\}/\{dQ(p,L)/dp)\}]_{p=p_{\rm c}}/\ln 2
\end{equation}
which is shown in Fig.~\ref{ytef} as a function of $L^{y_{\rm i}}$.

\subsubsection{Crossover of the effective magnetic exponent}
\label{yhcross}
\begin{figure}
\begin{center} \leavevmode \epsfxsize 8.6cm \epsfbox{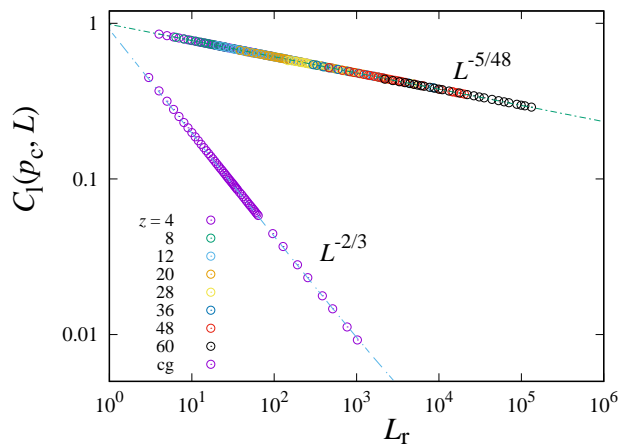}
\end{center}
\caption{Critical density of the largest cluster versus rescaled
finite-size $L_{\rm r}$. For the complete graph we define $L_{\rm r}=L$,
and for finite $z$, we define $L_{\rm r}$ as $L$ times an $z$-dependent
constant, chosen such that the large-$L$ data collapse in this figure.
Logarithmic scales are used in order to display the power-law nature
of these results. Straight lines are shown according to the exactly
known exponents. }
\label{larc}
\end{figure}

The determination of the magnetic exponent from the scaling behavior of
the density of the largest cluster, as described in Sec.~\ref{testu}
for the finite-$z$ models, was also done for the complete graph.
Figure \ref{larc} shows the combined results on logarithmic scales.
While the finite-$z$ data display a good data collapse for not too
small $L$, the results
for the complete graph fall on a line with a different slope.

We also quantitatively analyzed the scaling behavior of the density of
the largest cluster at the critical point $p_{\rm i,c}=1$.
Analogous to Eq.~(\ref{clfit}), it is expected to behave as
$c_{\rm l}(p_{\rm i,c},L) = g_0 L^{y_h-2} + \cdots  $.
Corresponding fits yielded the estimate $y_h = 1.3333$ (2), in a good
agreement with $y_h=4/3$ as given in Sec.~\ref{mfpxp} for the mean-field
model.

\begin{figure}
\begin{center} \leavevmode \epsfxsize 8.30cm \epsfbox{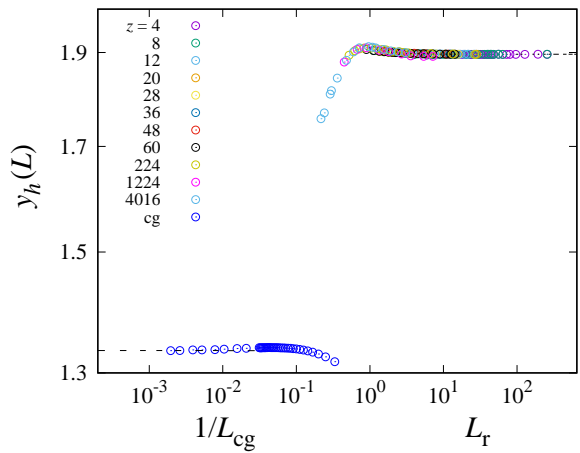}
\end{center}
\caption{Finite-size estimates of the magnetic exponent $y_h$ versus
rescaled finite size, as determined from the largest-cluster sizes
at the percolation threshold. }
\label{yhL}
\end{figure}

The exponent $y_h$ can be directly estimated by comparing two different
system sizes $L_1$ and $L_2$, as $y_h \approx
\ln[c_{\rm l}(p_{\rm i,c},L_1)/c_{\rm l}(p_{\rm i,c},L_2)]/\ln (L_1/L_2)+2$.
Figure \ref{yhL} displays the results for the models with finite $z$ as
well as for the complete graph.

\subsubsection{The crossover exponent at the MF fixed point}
\label{ycrmf}
\begin{figure}
\begin{center} \leavevmode 
\epsfxsize  8.0cm \epsfbox{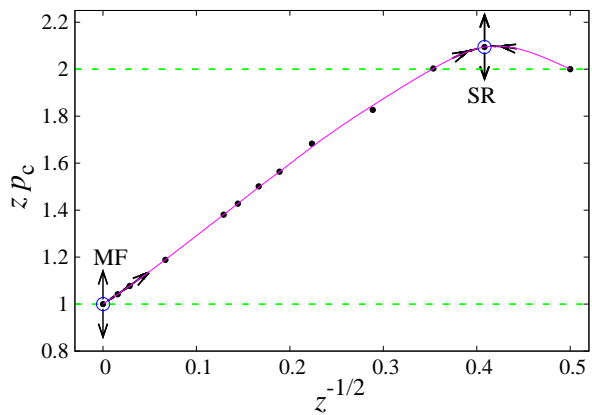}
\end{center}
\caption{Integrated critical bond probability $zp_{\rm c}$
versus $z^{-1/2}$ diagram. Shown are the
critical points (black circles) and a smoothed critical line on that basis.
The fixed points are shown as blue circles. The MF fixed
point is located  at $z=\infty$, and we have placed the short-range (SR)
fixed point at $z=6$, corresponding to the triangular lattice bond
percolation model.
Also shown is the renormalization flow near the fixed points. 
}
\label{pc}
\end{figure}

The critical points, in terms of the integrated bond probabilities
$p_{\rm i,c}=zp_{\rm c}$, are shown in Fig.~\ref{pc} versus $z^{-1/2}$.
The latter quantity describes the inverse range of the 
interactions when $z$ is large.
Among the renormalization exponents describing the flow in this
diagram, the one at the MF fixed point along the critical line is
still undetermined. This exponent, purportedly describing
the effect of interactions with a large but less than infinite range,
is denoted $y_{r}$. In order to determine this exponent, we
analyze the shape of the critical line $p_{\rm i,c}$.
For this purpose we parametrize the neighborhood of the MF fixed point
by a temperature variable  $v_t \equiv p_{\rm i}-1$, and an inverse range
parameter $v_r \equiv z^{-1/2}$. Under a renormalization transformation
by a scale factor $b$, these variables become
\begin{equation}
v_t(b)= b^{y_t}v_t \,~~~v_r(b)=b^{y_r}v_r
\label{rscal}
\end{equation}
It is essential that we allow that the inverse range renormalizes
with a new exponent $y_r$ that is different from the naive value 1.
It follows from Eq.~(\ref{rscal}) that the critical line in the
vicinity of the MF fixed point satisfies $v_t(b)\propto v_r(b)^{y_{t}/y_r}$,
so that $zp_{\rm c}-1 \propto (z^{-1/2})^{y_{t}/y_r}$
in Fig.~\ref{pc}.
The approximate linearity of the critical line near the MF fixed point
therefore suggests that $y_{r} =y_t=2/3$. This is further confirmed 
by a fit of $(a_1+a_2z^{-1/2})(z^{-1/2})^{1/\phi}$ to $zp_{\rm c}-1$,
for $z=224$, 1224 and 4016, using our Monte Carlo results
$p_{\rm c}=0.0053050415$ (33), 0.000880188 (90) and 0.0002594722 (11)
respectively.  This yielded a result for the crossover exponent as
$\phi^{-1}=y_{t}/y_r=1.006$ (7).

\section {Conclusion} 
\label{seccon} 
The numerical analyses of several models with coordination numbers up to
$z=60$ agree accurately with the universal constants $Q$, $R_{\rm b}$,
$R_{\rm e}$, and $R_{\rm 1}$. These models fit the theory of a short-range
fixed point describing all percolation models with a limited range of
interaction.

Near the short-range fixed point, the amplitudes of the corrections
to scaling were found to be approximately proportional to the square of 
the number of neighbors $z$, or the fourth power of the interaction
range $r$. These corrections are governed by the irrelevant exponent
$y_{\rm i}=-2$. Scaling then indicates that, in a vicinity of the
short-range fixed point, the interaction range scales as $r'=b^{-1/2}r$
under  a scale reduction with a factor $b$.

In this work, we have also included a few
models with larger $z$ up to $z=4016$, corresponding to a range $r=35.78$.
While the results of the numerical analysis of the large-$z$ models are
consistent with the universal constants mentioned, their accuracy falls
off when $z$ increases.
Using a rescaling of the finite-size parameter, a satisfactory data
collapse of the critical $Q$ versus $L$ curves could be obtained,
including the large $z$ models. These data illustrate the crossover of
$Q$ from the complete graph to the short-range universal value, as
illustrated in Fig.~\ref{crossover}.
 
The analysis of ${\rm d}Q/{\rm d}p$ in Sec.~\ref{crBr} indicates
that the models with finite $z$ behave in accordance with the short-range
temperature exponent $y_t=3/4$, again obeying short-range universality.
The behavior of the complete graph model is found to be different, with
$y_t=2/3$ instead, as predicted by the theory.
 
These results indicate that the crossover from MF to short-range
percolation takes place in a uniform way, with the MF limit
acting as a fixed point that is unstable along the critical line in 
the $p$ versus $z$ diagram, as illustrated in Fig.~\ref{pc}.
There is no sign of an intermediate higher critical fixed point along
this line. Such an additional fixed point would moreover imply that 
the MF fixed point is stable in the direction of the critical line,
contradicting the result $y_{r}=2/3$ in Sec.~\ref{ycrmf}.
The renormalization flow thus leads directly to the short-range
fixed point, which is stable in the range direction.
This situation is similar to that found
for the range-dependent crossover of the Ising model \cite{LBB}.


\appendix
\section{Analysis of the correction amplitudes}
\label{coramp}
\subsection{Wrapping probabilities}
\subsubsection{Almost spherical interaction regions}
The Monte Carlo data for the wrapping probabilities were subjected
to least-squares fits according to Eq.~(\ref{Rfit}). The results for the
amplitudes of the correction terms are listed in Tables~\ref{rb},
\ref{re} and \ref{r1}.
The residual $\chi^2$, compared with the number of degrees of freedom,
provides an indication of the reliability of the fits. One-sigma error
margins are  quoted, and they assume the validity and completeness of
Eq.~(\ref{Rfit}).

\begin{table*} 
\caption{Finite-size scaling of $R_{\rm b}$. The coordination number $z$,
the critical point $p_{\rm c}$, the  minimum system size $L_{\rm min}$
of the fits, and the amplitudes $b_{\rm i}$, $b_{\rm l}$ and $b_3$ of the
correction terms due to the irrelevant fields are listed.  }
\begin{tabular}
{|p{1.2cm}|p{2.5cm}|p{1cm}|p{2cm}|p{2cm}|p{2cm}|p{1.2cm}|p{1.2cm}|}
\hline $z$ & $p_{\rm c}$ & $L_{\rm min}$ & $b_{\rm i}$ & $b_{\rm l}$ &
$b_3$ & $\chi^2$ & $d_{\rm f}$ \\
\hline \multirow{2}*{4}
&  1/2       & 9  & $-$0.261(10)& $ $0.006(3) & $-$0.13(4) & 37.5& 40 \\[-.5ex]
&  1/2       &10  & $-$0.261(15)& 0.006(2)    & $-$0.13(6) & 37.1& 38 \\
\hline
\multirow{2}*{8}
& 0.25036843(9) & 8 & 0.34(2) &  $-$0.022(7)  & $-$0.28(7)  & 40.7& 47 \\
& 0.25036847(9) &10 & 0.31(5) &  $-$0.012(13) & $-$0.16(16) & 36.5& 45 \\
\hline
\multirow{2}*{12}
&0.15222023(4) & 7 & 0.39(1) &  $-$0.187(5)  &  $-$0.56(5) &74.5 & 73 \\[-.5ex]
&0.15222022(4) & 8 & 0.42(3) &  $-$0.194(9)  &  $-$0.7(1)  &73.6 & 72 \\
\hline
\multirow{2}*{20}
& 0.08415091(3) & 12 & 2.06(6) & $-$0.47(2) & $-$4.0(2)    & 30.1 & 45 \\[-.5ex]
& 0.08415093(3) & 13 & 1.98(8) & $-$0.45(2) & $-$3.7(4)    & 27.7 & 44 \\
\hline
\multirow{2}*{28}
& 0.05584923(3) & 12 & 5.2(1) & $-$1.16(3)  &  $-$14.2(4) & 18   & 26 \\[-.5ex]
& 0.05584925(3) & 14 & 5.0(2) & $-$1.11(4)  &  $-$13.3(7) & 15.5 & 25 \\
\hline
\multirow{2}*{36}   
&0.04169607(2) & 10 & 8.51(6)  & $-$1.89(2)  & $-$27.3(2)  & 68.6 & 60 \\[-.5ex] 
&0.04169608(2) & 11 & 8.40(8)  & $-$1.86(2)  & $-$26.8(3)  & 64   & 59 \\
\hline
\multirow{2}*{48}
&0.02974268(1) & 14 & 16.2(1)  &  $-$3.39(3) & $-$63.6(5)  & 46.2& 35 \\[-.5ex] 
&0.02974268(1) & 16 & 16.2(2)  &  $-$3.38(5) & $-$63(1)    & 45  & 33 \\
\hline
\multirow{2}*{60}
&0.02301189(2) & 16 & 24.2(8)  & $-$5.1(1)   & $-$98(4)    & 38.6& 37 \\[-.5ex]
&0.02301191(2) & 20 & 22.6(9)  & $-$4.5(2)   & $-$96(5)    & 36.6& 36 \\
\hline
\end{tabular}
\label{rb}
\end{table*}
\par The data in Tables~\ref{rb}, \ref{re} and \ref{r1} show that
the amplitudes $b_{\rm i}$ and $b_{\rm l}$ of the three wrapping
probabilities become very small between $z=4$ and $z=8$. This
is, for example, also visible in Fig.~\ref{rb48}.
\begin{table*}
\caption{Finite-size scaling analysis of $R_{\rm e}$. The
quantities listed are the same as those in Table~\ref{rb}.} \centering
\begin{tabular}
{|p{1.2cm}|p{2.5cm}|p{1cm}|p{2cm}|p{2cm}|p{2cm}|p{1.2cm}|p{1.2cm}|}
\hline $z$ & $p_{\rm c}$ & $L_{\rm min}$ & $b_{\rm i}$ & $b_{\rm l}$ &
$b_3$ & $\chi^2$ & $d_{\rm f}$ \\
\hline \multirow{2}*{4}
&  1/2          & 9 & 0.115(10) & 0.030(3)  & 0.15(3)  &  43  & 40   \\[-.5ex]
&  1/2          &10 & 0.120(14) & 0.029(4)  & 0.12(5)  &  43  & 38   \\
\hline \multirow{2}*{8}
& 0.25036863(10) &  9 & $-$0.11(4) &$-$0.009(11) &$ $0.28(12) & 61.1 & 41 \\
& 0.25036858(10) & 10 & $-$0.03(6) &$-$0.030(16) &$ $0.0(2)   & 57.7 & 40 \\
\hline \multirow{2}*{12}
& 0.15222030(7) & 10 & 0.64(7)   & $-$0.30(2) & 0.1(2)    & 57.1 & 65 \\[-.5ex]
& 0.1522204(1)  & 11 & 0.7(1)    & $-$0.30(3) & $-$0.1(5) & 52.2 & 64 \\
\hline \multirow{2}*{20}
& 0.08415101(3)  & 12 & 1.05(6) & $-$0.66(2) & $-$0.7(2)  & 36.3 & 51 \\[-.5ex]
& 0.08415102(3)  & 13 & 1.04(7) & $-$0.65(2) & $-$0.7(3)  & 36.2 & 50 \\
\hline \multirow{2}*{28}
& 0.05584935(4)  & 12 &  2.7(2)  & $-$1.45(4) & $-$2.5(7) & 14.7 & 28 \\[-.5ex]
& 0.05584933(5)  & 14 &  3.0(2)  & $-$1.51(6) & $-$4(1)   &   13 & 27 \\
\hline \multirow{2}*{36}
&0.04169613(2) &11& 4.8(1)       & $-$2.27(3) & $-$6.3(5) & 46.1 & 50 \\[-.5ex]
&0.04169613(2) &12& 4.9(2)       & $-$2.28(4) & $-$6.6(7) & 45.9 & 49 \\
\hline \multirow{2}*{48}
&0.02974271(1) &14& 10.1(2)      & $-$4.20(4) & $-$19.4(8)& 44   & 34 \\[-.5ex]
&0.02974271(1) &16& 10.2(3)      & $-$4.23(7) & $-$20(1)  & 41.6 & 32 \\ 
\hline \multirow{2}*{60}
&0.02301187(3) &16& 14.7(6)      & $-$5.9(1)  & $-$20(3)  & 30.8 & 41 \\[-.5ex]
&0.02301189(3) &20& 13(1)        & $-$5.6(2)  & $-$19(6)  & 28.1 & 40 \\
\hline
\end{tabular}
\label{re}
\end{table*}
\begin{table*}
\caption{Finite-size scaling analysis of $R_1$.
These data are not suitable for determining the critical
point $p_{\rm c}$ and we keep $p_{\rm c}$ in the fitting formula at fixed
values adopted from $R_{\rm b}$.} \centering
\begin{tabular}
{|p{1.2cm}|p{2.5cm}|p{1cm}|p{2cm}|p{2cm}|p{2cm}|p{1.2cm}|p{1.2cm}|}
\hline $z$ & $p_{\rm c}$ & $L_{\rm min}$ & $b_{\rm i}$ & $b_{\rm l}$ &
$b_3$ & $\chi^2$ & $d_{\rm f}$ \\
\hline \multirow{2}*{4}
&  1/2      &  9 &   0.188(5)  & 0.012(1)  &  0.14(2)  & 41   &  40 \\[-.5ex]
&  1/2      & 10 &   0.189(7)  & 0.012(2)  &  0.13(3)  & 40.8 &  38 \\
\hline \multirow{2}*{8}
& 0.2503685 & 9 & $-$0.184(17) & $-$0.005(2) & 0.14(6)  & 66   & 46  \\[-.5ex]
& 0.2503685 &10 & $-$0.145(24) & $-$0.016(7) & 0.00(9)  & 62   & 45  \\
\hline \multirow{2}*{12}
& 0.1522203  & 10 &  0.17(2)  &  $-$0.069(6) & 0.1(1)  & 68.7  & 71  \\[-.5ex]
& 0.1522203  & 11 &  0.19(3)  &  $-$0.073(7) & 0.0(1)  & 68.2  & 70  \\
\hline \multirow{2}*{20}
& 0.0841509  & 12 & $-$0.44(2) & $-$0.111(7) & 1.4(1)  & 47.9  & 60 \\[-.5ex]
& 0.0841509  & 13 & $-$0.41(3) & $-$0.119(8) & 1.3(1)  & 44.6  & 59 \\
\hline \multirow{2}*{28}
& 0.0558493  & 12 & $-$1.10(6) & $-$0.18(2) & 5.4(3)  & 24.4  & 29  \\[-.5ex]
& 0.0558493  & 14 & $-$0.91(9) & $-$0.23(2) & 4.3(4)  & 15.2  & 28   \\
\hline \multirow{2}*{36} 
& 0.04169608  & 11 & $-$1.72(5) & $-$0.22(1) & 10.0(2) & 63.4  & 61 \\[-.5ex]
& 0.04169608  & 12 & $-$1.63(7) & $-$0.25(2) & 9.6(3)  & 59.3  & 60 \\
\hline \multirow{2}*{48} 
& 0.02974268  & 12 & $-$2.95(5) & $-$0.43(1) & 21.7(2) & 45.2 & 36 \\[-.5ex]
& 0.02974268  & 14 & $-$2.95(7) & $-$0.43(2) & 21.7(3) & 44.5 & 34 \\
\hline \multirow{2}*{60}
&  0.0230119  & 16 & $-$5.2(2) & $-$0.43(5) & 43(1)  & 19.6  &  37 \\[-.5ex]
&  0.0230119  & 20 & $-$5.3(5) & $-$0.4(1)  & 43(3)  & 19.5  &  36 \\
\hline
\end{tabular}
\label{r1}
\end{table*}
All three amplitudes grow significantly in absolute value with $z>8$,
in accordance with the interpretation of $z$ as a
measure of the distance to the short-range fixed point. An analysis
of  $b_{\rm i}(z)$ and $b_{\rm l}(z)$ reveals a linear dependence
on $z^2$ for systems with ``almost circular'' interaction regions.
This behavior was illustrated in Fig.~\ref{bli}.

\subsubsection{Range dependence and fourfold symmetry perturbation}
\label{4fold}
\begin{figure} 
\begin{center} \leavevmode \epsfxsize  6.0cm \epsfbox{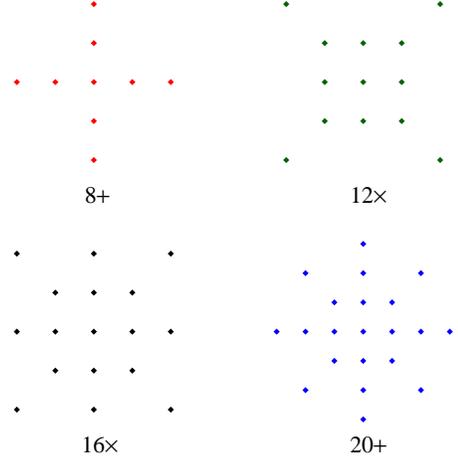}
\end{center} 
\caption{Positions of the neighbors interacting with the
central site. These four subfigures define models that disobey the
condition that the interacting neighbors fill a circle.}
\label{posi} 
\end{figure}

While the local interactions assume an ``almost circular'' form
for the large-$z$ models investigated thus far, the deviations from
isotropy are obvious for small $z$. To study the possible effects of 
these 4-fold deviations from circular symmetry on the correction
amplitudes, we introduce a few models that do not obey the rule that
the set of equivalent neighbors fills a circle with a given range $r$.
Analysis of these models could possibly distinguish between the effects
of the coordination number $z$ and the distances to these neighbors.
We refer to the newly chosen models as the 8+ model, the 12$\times$
model, the 16$\times$ model, and the 20+ model.  The ``$\times$'' and
``+'' symbols show the orientation of the fourfold perturbation.  These
models are defined in Fig.~\ref{posi}. We do not consider the ``$4\times$''
model with interacting neighbors at $(x,y)=(\pm1,\pm1)$, which decouples
into two disjoint percolation models, and thus belongs to a different
category.

\begin{table*}
\caption{Finite-size-scaling analysis of $R_{\rm b}$ for several models
that do not have ``almost circular'' local interactions.
The table includes the coordination number $z$, the critical point
$p_{\rm c}$, the correction amplitudes $b_{\rm i}$ and $b_{\rm l}$, 
and the amplitude $b_3$ of the $L^{-3}$ correction term. These correction
amplitudes are to be compared with those for the ``almost circular'' models
in Table~\ref{rb}.}
\begin{tabular}
{|p{1.2cm}|p{2.5cm}|p{1cm}|p{2cm}|p{2cm}|p{2cm}|p{1.2cm}|p{1.2cm}|}
\hline $z$ & $p_{\rm c}$ & $L_{\rm min}$ & $b_{\rm i}$ & $b_{\rm l}$ &
$b_3$ & $\chi^2$ & $d_{\rm f}$ \\
\hline \multirow{2}*{8+} &
0.22149934(6) &8 & 0.82(3)&$-$0.364(9) &$-$1.8(1) &60.1 &64  \\[-.5ex] &
0.22149937(6)  &9 & 0.77(4)& $-$0.35(1)  & $-$1.7(2)  &  57  &63
\\
\hline

\multirow{2}*{12$\times$}  & 0.13805337(9) &11 &4.9(2) & $-$0.79(4)  &$-$13.1(6)
&  49 & 50\\[-.5ex]
& 0.13805340(9) &12 & 4.8(2)&  $-$0.75(5) &$-$12(1)  & 47.6 & 49 \\
\hline

\multirow{2}*{16$\times$} & 0.10321771(9) & 9 & 3.9(1) & $-$0.70(3)& $-$9.8(4)
& 40& 48 \\[-.5ex] 
& 0.10321772(9) & 10 & 3.9(1) & $-$0.69(4) & $-$9.6(5)  & 39.6  & 47\\
\hline

\multirow{2}*{20$+$} &0.07831101(9) & 11&6.0(2) & $-$1.38(5) & $-$17.4(7)
&39.4 &44 \\[-.5ex]
& 0.07831101(9)  & 12 & 6.0(2)&$-$1.39(6) & $-$17.4(9) & 36.1&40
\\
\hline
\end{tabular} \label{4-fold-b}
\end{table*}
\begin{table*}
\caption{Finite-size scaling analysis of $R_{\rm e}$ for  several models 
that do not have ``almost circular'' local interactions.
The coordination number $z$, critical point
$p_{\rm c}$, degenerate irrelevant field amplitudes $b_{\rm i}$ and
$b_{\rm l}$, amplitude $b_3$ of $L^{-3}$ correction term are listed.}

\begin{tabular}
{|p{1.2cm}|p{2.5cm}|p{1cm}|p{2cm}|p{2cm}|p{2cm}|p{1.2cm}|p{1.2cm}|}
\hline $z$ & $p_{\rm c}$ & $L_{\rm min}$ & $b_{\rm i}$ & $b_{\rm l}$ &
$b_3$ & $\chi^2$ & $d_{\rm f}$ \\
\hline

\multirow{2}*{8+} &  0.22149959(7)  &      11   &    0.78(9) &   $-$0.41(3) &
2.4(4)     &    62.9  & 57       \\[-.5ex] &   0.22149957(7)   &     12 &
0.9(1)   &   $-$0.43(3)    &    2.0(4)      & 61   &56   \\
\hline

\multirow{2}*{12$\times$} &     0.1380535(1)  &  11&  1.2(1)   &  $-$1.02(3) &
$-$1.5(5) &  69.4   & 79 \\[-.5ex] &0.1380537(1)    &  12  &   0.7(2)     &
$-$0.89(6)    &  1(1)   &  61.8  &  78    \\
\hline

\multirow{2}*{16$\times$} & 0.1032180(1)&12 & 1.3(2) & $-$0.96(6) & $-$1.1(9) &
48.4 & 43 \\[-.5ex] & 0.1032179(1) & 14  &  1.6(3)  &  $-$1.04(8)    & $-$3(1)
& 45.5  &  40   \\
\hline

\multirow{2}*{20$+$} & 0.07831096(9)  & 11 & 3.1(2)  & $-$1.57(5) &$-$2.4(8)
& 47.6 &  44\\[-.5ex] & 0.07831095(9)  & 12& 3.2(2) & $-$1.59(6) &$-$2.8(9) 
& 46.7 &42  \\
\hline
\end{tabular} \label{4-fold-e}
\end{table*}
\begin{table*}
\caption{Analysis of the correction amplitudes in $R_1$ for  several
models that do not have ``almost circular'' local interactions.}
\centering
\begin{tabular}
{|p{1.2cm}|p{2.5cm}|p{1cm}|p{2cm}|p{2cm}|p{2cm}|p{1.2cm}|p{1.2cm}|}
\hline $z$ & $p_{\rm c}$ & $L_{\rm min}$ & $b_{\rm i}$ & $b_{\rm l}$ &
$b_3$ & $\chi^2$ & $d_{\rm f}$ \\
\hline \multirow{2}*{8+} &
0.2214995      &    11  &     0.11(4)   &  $-$0.06(1)     &  1.6(2)    &
60   &  58      \\[-.5ex] &   0.2214995      & 12    &    0.14(4)   &
$-$0.07(1)     &   1.5(2) & 57.5  &  57    \\
\hline
\multirow{2}*{12$\times$} &  0.13805374   &  11& $-$1.80(5)  & $-$0.14(1)
& 5.4(2)  & 67 & 55 \\[-.5ex] & 0.13805374    &   12 &  $-$1.87(7)
&$-$0.12(2)  & 5.9(4)&65.1  &54  \\
\hline
\multirow{2}*{16$\times$} &0.1032177 & 14 & $-$1.1(1)  & $-$0.17(3) &3.3(6)
& 37 & 37
\\[-.5ex] &0.1032177 & 16 & $-$0.9(2)  & $-$0.22(4) &1.9(9) & 31.3& 34 \\
\hline

\multirow{2}*{20$+$} & 0.0783110 & 11 & $-$1.40(7)  & $-$0.10(2) &7.3(3)  &
47.3 & 46 \\[-.5ex] &0.0783110 & 12 & $-$1.41(9)  & $-$0.10(2) &7.4(4) & 45.9
& 42
\\
\hline
\end{tabular}
\end{table*}

\begin{figure}
\begin{center}
\leavevmode \epsfxsize 21cm \epsfbox{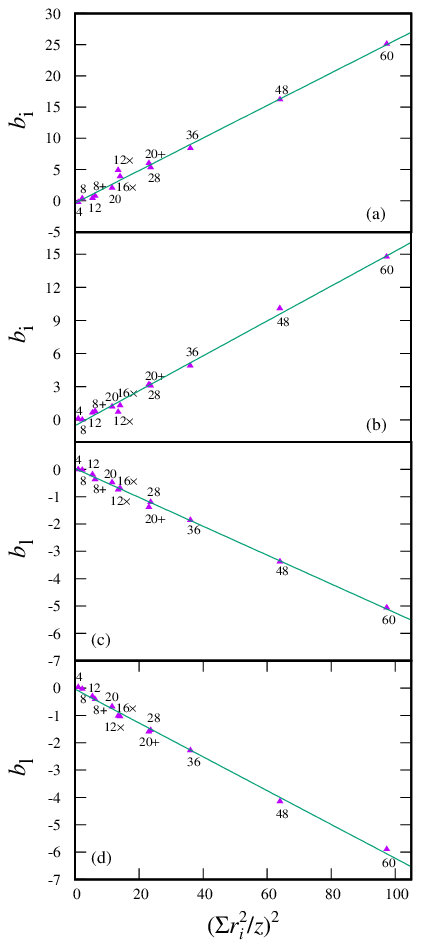}
\vspace{1em}
\end{center}
\caption{Correction amplitudes in $R_{\rm b}$ and $R_{\rm e}$ vs.~ the
square of the sum $\Sigma r_{i}^2/z$. (a): $b_{\rm i}$ in $R_{\rm b}$;
(b): $b_{\rm i}$ in $R_{\rm e}$; (c): $b_{\rm l}$ in $R_{\rm b}$;
(d): $b_{\rm l}$ in $R_{\rm e}$. The data points in the figures are
labeled with the corresponding coordination numbers $z$.}
\label{multib}
\end{figure}

Monte Carlo simulations, and numerical analyses of the wrapping probabilities
were performed, similar to those of the ``almost circular'' cases.
The results are listed in Table \ref{4-fold-b}. We found that the correction
amplitudes did not fit in the simple picture of Fig.~\ref{bli}, i.e.,
neither $b_{\rm i}$ nor $b_{\rm l}$ can be simply described by a quadratic
function of $z$. To allow for the distances spanned by the $z$ interactions,
we considered the dependence of the amplitudes on the sums $\sum r_i^kz$,
for several $k$, and on powers of $z$.
As shown in Figs.~\ref{multib}, the amplitudes $b_{\rm i}$
and $b_{\rm l}$ in $R_{\rm b}$ and $R_{\rm e}$ behave almost linearly as
a function of $(\sum r_i^2/z)^2$. 

We have attempted to further improve the description of the dependence 
of $b_{\rm i}$ and $b_{\rm l}$  on the local interaction neighborhood, 
by including a few more
sums over the interacting neighbors, involving polar coordinates $r_i$ and 
the angle $\phi_i$ with respect to a nearest neighbor.
While a fit of the data by
\begin{displaymath}
b_{\rm i/l} =
c_0(\sum_i r_i / z)^2+c_1\sum_i
r_i^2+c_2\sum_i r_i\cos(4\phi_i)+
\mbox{\hspace{35mm}}
\end{displaymath}
\begin{equation}
c_3[\sum_i r_i\cos(4\phi_i)]^2
+c_4 \sum_i r_i^2/z +c_5(\sum_i r_i^2/z)^2+c_6 \sum_i r_i^4 \,
\label{il_be}
\end{equation}
did reduce the residual $\chi^2$, we doubt the
physical meaning of the resulting parameters, which were seen to
depend considerably on variations of the fit formula. No clear 
dependence of the amplitudes on the orientation of the anisotropy, i.e.,
on the presence of terms with $\cos(4\phi_i)$, was found.

\begin{table*}
\caption{Finite-size amplitudes of the critical Binder ratio $Q$,
for several values of the coordination number $z$. For finite $z$, fits
were applied according to Eq.~(\ref{fsr}) with an additional term
$c_1 (p-p_{\rm c})L^{y_t-2}$, which accounts for the dependence of the
$p$-dependent term on the irrelevant field.
For $z>4$, $Q$ was
fixed at the universal value determined for the 4-neighbor model,
and the percolation threshold $p_{\rm c}$ was fixed at the value found
earlier from the wrapping probabilities. The errors between parentheses
are shown as one-sigma margins in the last decimal place listed.
Actual uncertainties can be
a few times larger after taking into account the errors in $p_{\rm c}$
and $Q$.  The exponents $y_t$ and $y_h$ were fixed at their
exactly known values in all cases. 
For the complete graph (cg) data, the value of $Q$ was left free in the
fit formula, while the percolation threshold was fixed at the integrated
probability $p_{\rm i,c}=1$.  Three finite-size correction terms,
with fixed exponents $-2/3$, $-4/3$ and $-2$ were included for the
complete graph case. The respective amplitudes are given in the last
three columns.}
\begin{tabular}{|p{0.8cm}|p{1cm}|p{2.0cm}|p{2.2cm}|p{2.5cm}|p{2.3cm}|p{2cm}|}
\hline
$z$  & $L_{\rm min}$ & $p_{\rm c}$ &$Q$&$b_{\rm i}$ &$b_{\rm l}$ &$b_3$ \\
\hline \multirow{2}*{4}   
& 10   &   1/2       & 0.87057(1)&    3.68(15)   & 1.25(12)   & $-$4.1(3) \\
& 11   &             & 0.87057(1)&    3.79(19)   & 1.34(14)   & $-$4.3(4) \\
\hline \multirow{2}*{8}
& 13  &   0.2503685  &  0.87057  &   0.37(15)    & $-$1.0(2)  &$ $0.9(4) \\
& 14  &              &           &   0.42(17)    & $-$0.9(2)  & 0.7(5)   \\
\hline \multirow{2}*{12}
& 16  & 0.1522203    & 0.87057   &  $-$8.5(2)    & $-$9.7(4)  & 18.2(8)  \\
& 18  &              &           &  $-$8.4(2)    & $-$9.4(5)  & 17.6(10) \\
\hline \multirow{2}*{20}   
& 18  & 0.0841509    & 0.87057   & $-$17.9(3)    & $-$22(1)   & 42(2)    \\
& 20  &              &           & $-$17.8(4)    & $-$22(1)   & 42(3)    \\
\hline \multirow{2}*{28}
&  20 & 0.0558493    & 0.87057   & $-$29.9(2)    & $-$48.8(5) & 88(1)   \\
&  22 &              &           & $-$29.9(2)    & $-$49.1(7) & 89(1)   \\
\hline \multirow{2}*{36}
& 22  & 0.04169608   & 0.87057   & $-$35.8(2)    & $-$70.9(8) & 125(2)  \\
& 24  &              &           & $-$36.0(2)    & $-$69.9(9) & 124(2)  \\
\hline \multirow{2}*{48}
& 24  & 0.02974268   & 0.87057   & $-$37.2(2)    & $-$121(1)  & 206(1)  \\
& 28  &              &           & $-$37.7(3)    & $-$119(1)  & 203(2)  \\
\hline \multirow{2}*{60}
& 24  & 0.0230119    & 0.87057   & $-$31.1(10)   & $-$184(5)  & 307(9)   \\
& 28  &              &           & $-$29.9(14)   & $-$190(7)  & 316(12)  \\
\hline \hline \multirow{2}*{\rm cg}
&  19 & 1           &0.38127(5) &  0.158(2)     & 0.13(3)    & 0.25(10)  \\
&  20 &             &0.38125(6) &  0.159(2)     & 0.12(3)    & 0.32(11)  \\
\hline
\end{tabular} \label{qtab}
\end{table*}
\subsection{Correction amplitudes of the Binder ratio}
Least-squares fits according to Eq.~(\ref{fsq}) to the finite-size data for the
Binder ratio gave clear signs of the presence of the term $b_4 L^{y_t-2y_h}$, i.e.,
the existence of a quadratic contribution of the physical magnetic field to the
temperature field.  Results these fits  are listed in Table \ref{qtab}.
In particular the data for $b_{\rm i}$, which were also shown in
Fig.~\ref{bizq}, display a different behavior than, for instance,
the approximately linear dependence on $z^2$ as seen in Fig.~\ref{bli}.
This suggests that the finite-size-scaling functions of the magnetization
moments depend nonlinearly on the irrelevant field associated with the 
interaction range.

\acknowledgments
Y.~D. thanks the Ministry of Science and Technology of China for Grant
No. 2016YFA0301604 and the National Natural Science Foundation of China
for Grant  No. 11625522.
Y.~O. acknowledges the hospitality of the Lorentz Institute of the Leiden
University, and  H.~B. thanks the University of Science and
Technology of China in Hefei for hospitality extended to him.

\end{document}